\documentclass[ALICE,manyauthors]{cernphprep}

\usepackage[comma,square,numbers,sort&compress]{natbib}
\usepackage{hyperref}
\usepackage{lineno}
\usepackage[bottom]{footmisc}

\newcommand{\kzs}{\rm K^0_S}

\newcommand{\pvec}[1]{\vec{#1}\mkern2mu\vphantom{#1}}
\newcommand{\kstar}{k^*}
\newcommand{\vkstar}{\pvec{k}^*}
\newcommand{\rstar}{r^*}
\newcommand{\vrstar}{\pvec{r}^*}

\usepackage{hyperref}
\usepackage{multirow}

\begin{document}%

%%%%%%%%%%%%%%%  Title page %%%%%%%%%%%%%%%%%%%%%%%%
\begin{titlepage}
\PHyear{2018}
\PHnumber{234}      % required, will be obtained from PH
\PHdate{29 August}  % required, will be obtained from PH
%\PHdate{\today}
%

%%% Put your own title + short title here:
\title{Measuring K$^0_{\rm S}$K$^{\rm \pm}$ interactions \\
using pp collisions at $\mathbf {\sqrt{s}=7}$ TeV}
\ShortTitle{K$^0_{\rm S}$K$^{\rm \pm}$ interaction using
pp collisions}   % appears on right page headers

%%% Do not change the next lines
\Collaboration{ALICE Collaboration\thanks{See Appendix~\ref{app:collab} for the list of collaboration members}}
\ShortAuthor{ALICE Collaboration} % appears on left page headers, do not change

\begin{abstract}
We present the first measurements of femtoscopic correlations between the K$^0_{\rm S}$ and K$^{\rm \pm}$ particles in pp collisions at $\sqrt{s}=7$ TeV measured by the ALICE experiment.  The observed femtoscopic correlations are consistent with final-state interactions proceeding solely via the $a_0(980)$ resonance. 
The extracted kaon source radius and correlation strength parameters for K$^0_{\rm S}$K$^{\rm -}$ are found to be equal
within the experimental uncertainties to those for K$^0_{\rm S}$K$^{\rm +}$.
Results of the present study are compared with those
from identical-kaon femtoscopic studies also performed with pp collisions at $\sqrt{s}=7$ TeV by ALICE
and with a K$^0_{\rm S}$K$^{\rm \pm}$ measurement in 
Pb-Pb collisions at $\sqrt{s_{\rm NN}}=2.76$ TeV.
Combined with the Pb-Pb results, our pp analysis is found to be compatible with the interpretation of 
the $a_0(980)$ having a tetraquark structure instead of that of a diquark.
\end{abstract}
\end{titlepage}
\setcounter{page}{2}

\section{Introduction}
Recently, by using Pb-Pb collisions at $\sqrt{s_{\rm NN}}=2.76$ TeV, the ALICE 
experiment~\cite{Aamodt:2008zz} has published the first-ever study of K$^0_{\rm S}$K$^{\rm \pm}$
femtoscopy~\cite{Acharya:2017jks}. K$^0_{\rm S}$K$^{\rm \pm}$
femtoscopy differs from identical-kaon femtoscopy, for which a number of studies exist in the 
literature~\cite{Abelev:2006gu,Abelev:2012ms,Abelev:2012sq,Adam:2015vja}, in that the only pair interaction expected is a final-state interaction (FSI) through the
$a_0(980)$ resonance. It was found in that Pb-Pb study that the FSI in K$^0_{\rm S}$K$^{\rm \pm}$
proceeds solely through the $a_0(980)$ resonance, i.e. with no competing non-resonant channels, 
and the extracted kaon source parameters
agree with published results from identical-kaon studies in Pb-Pb collisions. These results were 
found to be compatible with the interpretation of the $a_0$ resonance 
as a tetraquark state rather than a diquark$^1$\footnotetext[1]{Note that the term ``diquark'' will be used in this paper to indicate a
$q_i{\overline q_j}$ quark pair.} state~\cite{Acharya:2017jks,Achasov:2002ir,Santopinto:2006my,Jaffe:1976ig}. A recent theoretical calculation has shown that 
the ALICE Pb-Pb results can indeed be described by a model based on the four-quark 
model~\cite{Achasov:2017zhy}.

The argument given in Ref.\cite{Acharya:2017jks} for a tetraquark 
$a_0$ being compatible with the Pb-Pb 
K$^0_{\rm S}$K$^{\rm \pm}$ result stated above is
based on two factors: 1) the kaon source geometry, and 2) an empirical selection rule
(For the sake of simplicity of notation, ``$a_0$'' will be used for the remainder of this
paper to represent ``$a_0(980)$.'').
For factor 1), the production cross section 
of the $a_0$ resonance in a reaction channel such as 
${\rm K}^0{\rm K}^-\rightarrow a^-_0$ should depend on whether the $a^-_0$ is 
composed of ${\rm d\overline{u}}$ or ${\rm d\overline{s}s\overline{u}}$ quarks, the 
former requiring the annihilation of the ${\rm \overline{s}s}$ pair and the latter being 
a direct transfer of the valence quarks from the kaons to the $a^-_0$. Since the femtoscopic size of
the 0-10\% most central Pb-Pb collision is measured to be 5--6 fm, the large geometry
in these collisions is favorable for the direct transfer of quarks to the $a_0$, whereas
not favorable for the annihilation of the strange quarks due to the short-ranged nature of
the strong interaction. For factor 2), the direct transfer of the valence quarks from the kaons to the 
$a^-_0$ is favored since this is an ``OZI superallowed''  reaction~\cite{Jaffe:1976ig}. 
The OZI rule can be stated as ``an inhibition associated with the creation or annihilation 
of quark lines''~\cite{Jaffe:1976ig}. Thus, the annihilation of the strange quarks is suppressed
by the OZI rule. Both of these factors favor the formation of a
tetraquark $a_0$ and suppress the formation of a diquark $a_0$. As a result of
this, if the $a_0$ were a diquark, one would expect competing non-resonant channels
present and/or no FSI at all, i.e. free-streaming, of the kaon pair thus diluting the strength of the 
$a_0$ resonant FSI. The fact that this is not seen to be the case in Pb-Pb collisions 
favors the tetraquark $a_0$ interpretation. 

The geometry of the kaon source is seen to be an important factor in the argument given above,
i.e. the large kaon source seen in Pb-Pb collisions suppresses the annihilation of the 
strange quarks in the kaon pair
and enhances the direct transfer of quarks to the $a_0$. It is interesting to speculate
on the dependence of the strength of the $a_0$ resonant FSI on the size of the kaon source,
particularly for a very small source of size $\sim 1$ fm that would be obtained in 
pp collisions~\cite{Abelev:2012ms,Abelev:2012sq}.
For a kaon source of size $\sim 1$ fm, the kaons in a produced kaon pair would be 
overlapping with each other at the source,
thus giving a geometric enhancement of the strange-quark annihilation channel that could
compete with, or even dominate over, the OZI rule suppression of quark annihilation. Thus we might expect
that the tetraquark $a_0$ resonant FSI could be diluted or completely suppressed by competing non-resonant annihilation
channels that could open up, whereas a diquark $a_0$ resonant FSI, which was not seen to be suppressed
by either geometry or the OZI rule in Pb-Pb, would not be diluted. 
A femtoscopic measurement of 
K$^0_{\rm S}$K$^{\rm \pm}$ correlations in pp collisions should be able to test this
by determining
the strength of the $a_0$ FSI by measuring the femtoscopic $\lambda$ parameter.
In more concrete terms, if we were to compare the $\lambda$ parameters extracted in
K$^0_{\rm S}$K$^{\rm \pm}$ femtoscopic measurements in pp collisions and Pb-Pb
collisions, for a tetraquark $a_0$ we would 
expect $\lambda_{K^0_SK^\pm ({\rm PbPb})}>\lambda_{K^0_SK^\pm ({\rm pp})}$
whereas for a diquark $a_0$ we would expect 
$\lambda_{K^0_SK^\pm ({\rm PbPb})}\sim\lambda_{K^0_SK^\pm ({\rm pp})}$.
An independent check could also be made by comparing $\lambda$ 
from K$^0_{\rm S}$K$^{\rm \pm}$ femtoscopy in pp collisions with $\lambda$ from identical-kaon
femtoscopy in pp collisions in a similar way as was done for Pb-Pb collisions~\cite{Acharya:2017jks}. 
Since we expect identical-kaon correlations to go solely through quantum statistics (and FSI
for neutral kaons), our expectation for a tetraquark $a_0$ would be 
$\lambda_{KK({\rm pp})}>\lambda_{K^0_SK^\pm({\rm pp})}$
whereas for a diquark $a_0$ we would expect $\lambda_{KK({\rm pp})}\sim\lambda_{K^0_SK^\pm({\rm pp})}$.

In this Letter, femtoscopic correlations with the particle pair combinations K$^0_{\rm S}$K$^{\rm \pm}$ are studied for the first time in pp collisions at $\sqrt{s}=7$ TeV by the ALICE experiment.  
The physics goals of the present K$^0_{\rm S}$K$^{\rm \pm}$ femtoscopy study are the following:
1) show to what extent the FSI through the $a_0$ resonance describes the correlation functions,
2) study the K$^0$ and ${\rm \overline{K}^0}$ sources to see if there are differences in the source parameters,
3) compare the results of the extracted kaon source parameters from the present study with
the published results from Pb-Pb collisions and identical kaon results from pp collisions, and
4) see if the results from this pp study are compatible with a tetraquark $a_0$ as suggested
from the Pb-Pb study.

\section{Description of experiment and data selection}
The ALICE experiment and its performance in the LHC Run 1 $(2009-2013)$ are described in 
Ref.~\cite{Aamodt:2008zz} and Ref.~\cite{Abelev:2014ffa,Akindinov:2013tea}, respectively.
About $370\times 10^6$ minimum-bias 7 TeV pp collision events taken in 2010 were used in this analysis.
Events were classified using the measured amplitudes in the 
V0 detectors, which consist of two arrays of scintillators located along the beamline and 
covering the full
azimuth~\cite{Abelev:2013vea,Abelev:2013qoq}. 
Charged particles were reconstructed and identified with the central barrel detectors located 
within a solenoid magnet with a field strength of $B=\pm 0.5$ T.
Charged particle tracking was performed using the 
Time Projection Chamber (TPC)~\cite{Alme:2010ke} 
and the Inner Tracking System (ITS) \cite{Aamodt:2008zz}. The ITS allowed for high spatial resolution in determining the primary (collision) vertex.
A momentum resolution of less than 10 MeV/$c$ was typically obtained for the charged tracks
of interest in this analysis~\cite{Alessandro:2006yt}.
The primary vertex was obtained
from the ITS, the position of the primary vertex being constrained along the beam direction (the ``$z$-position'') to be within $\pm10$ cm of the center of the ALICE detector.
In addition to the standard track quality selections~\cite{Alessandro:2006yt}, the selections based on the quality of track fitting and the number of detected tracking points in the TPC were used to ensure that only well-reconstructed tracks were taken in the analysis~\cite{Abelev:2014ffa,Alme:2010ke,Alessandro:2006yt}.

Particle identification (PID) for reconstructed tracks was carried out using both the TPC and the Time-of-Flight (TOF) detectors in the pseudorapidity range $|\eta| < 0.8$~\cite{Abelev:2014ffa,Akindinov:2013tea}.
For the PID signal from both detectors, a value was assigned to each track denoting the number of standard deviations between the measured track information and calculated values ($N_{\sigma}$) \cite{Adam:2015vja,Abelev:2014ffa,Akindinov:2013tea,Alessandro:2006yt}.
For TPC PID, a parametrized Bethe-Bloch formula was used to calculate the specific energy 
loss $\left<{\rm d}E/{\rm d}x\right>$ in the detector expected for a particle with a given mass and momentum. For PID with TOF, the particle mass was used to calculate the expected time-of-flight as a function of track length and momentum. 
This procedure was repeated for four ``particle species hypotheses'', i.e. electron, pion, kaon and proton, and, for each hypothesis, a different $N_{\sigma}$ value was obtained per detector.

\subsection{Kaon selection}
The methods used to select and identify individual K$^0_{\rm S}$ and K$^{\rm \pm}$ particles are the same as those used for the ALICE K$^0_{\rm S}$K$^0_{\rm S}$~\cite{Abelev:2012ms}
and K$^{\rm \pm}$K$^{\rm \pm}$~\cite{Abelev:2012sq} analyses from $\sqrt{s}=7$ TeV pp collisions.
These are now described below.

\subsubsection{K$^0_{\rm S}$ selection}
The K$^0_{\rm S}$ particles were reconstructed from the decay K$^0_{\rm S}\rightarrow\pi^+\pi^-$, with the daughter $\pi^+$ and $\pi^-$ tracks detected in the TPC, ITS and TOF detectors.
The secondary vertex finder used to locate the neutral kaon decays employed the ``on-the-fly'' reconstruction method~\cite{Alessandro:2006yt}, which recalculates the daughter track momenta during the original tracking process under the assumption that the tracks came from a decay vertex instead of the primary vertex.
Pions with $p_{\rm T}>0.15$ GeV/$c$ were accepted
(since for lower $p_{\rm T}$ track finding efficiency drops rapidly)
and the distance of closest approach to the primary vertex (DCA) of the reconstructed K$^0_{\rm S}$ was required to be less than 0.3 cm in all directions. 
The required $N_{\sigma}$ values for the pions were $N_\sigma^{\rm TPC} < 3$ (for all momenta) and
$N_\sigma^{\rm TOF} < 3$ for $p>0.8$ GeV/$c$. An invariant mass distribution for the $\pi^+\pi^-$
pairs was produced and the K$^0_{\rm S}$ was defined to be resulting from a pair that fell
into the invariant mass range $0.480<m_{\pi^+\pi^-}<0.515$ GeV/$c^2$, corresponding to
$\pm4.7\sigma$, where $\sigma=3.7$ MeV/$c^2$ is the width of a Gaussian fit to the invariant mass distribution.

\subsubsection{K$^\pm$ selection}
Charged kaon tracks were detected using the TPC and TOF detectors,
and were accepted if they were within the range 
$0.14<p_{\rm T}<1.2$ GeV/$c$ in order to obtain good PID.
The determination of the momenta of the tracks was performed using tracks reconstructed with the TPC only and constrained to the primary vertex.
In order to reduce the number of secondary tracks (for instance the charged particles produced in the detector material, particles from weak decays, etc.),
the primary charged kaon tracks were selected based on the DCA, such that the DCA
transverse to the beam direction was less than 2.4 cm and the DCA along the beam direction was
less than 3.2 cm. 
If the TOF signal were not available, 
the required $N_{\sigma}$ values for the charged kaons were $N_\sigma^{\rm TPC} < 2$ for $p_{\rm T}<0.5$ GeV/$c$, and the track was rejected for $p_{\rm T}>0.5$ GeV/$c$. If the TOF signal were also available and $p_{\rm T}>0.5$ GeV/$c$: $N_\sigma^{\rm TPC} < 2$
and $N_\sigma^{\rm TOF} < 2$ ($0.5<p_{\rm T}<1.2$ GeV/$c$).

The K$^0_{\rm S}$K$^{\rm \pm}$ experimental pair purity was estimated from a Monte Carlo (MC) study based on PYTHIA~\cite{Sjostrand:2006za} simulations with the 
Perugia2011 tune~\cite{Skands:2010ak}, and
using GEANT3~\cite{Brun:1994aa}  to model particle transport through the ALICE detectors. The purity
was determined from the fraction of the reconstructed MC simulated pairs that were identified as known K$^0_{\rm S}$K$^{\rm \pm}$ pairs from PYTHIA. The pair purity was
estimated to be $\sim 83$\% for all kinematic regions studied in this analysis.
The single-particle purities for K$^0_{\rm S}$ and K$^{\rm \pm}$ particles used in this analysis were
estimated to be $\sim 92$\% and $\sim 91$\%, respectively. The uncertainty in calculating the pair
purity is estimated to be $\pm 1$\%.

\section{Analysis methods}

\subsection{Experimental Correlation Functions}
This analysis studies the momentum correlations of K$^0_{\rm S}$K$^{\rm \pm}$ pairs using the two-particle correlation function, defined as 

\begin{equation}
C(k^*)=\frac{A(k^*)}{B(k^*)},
\label{Cexp}
\end{equation}

where $A(k^*)$ is the measured distribution of pairs from the same event, $B(k^*)$ is the reference distribution of pairs from mixed events,
and $k^*$ is the magnitude of the momentum of each of the particles in the pair rest frame (PRF),

\begin{equation}
k^*=\sqrt{\frac{(s-m_{\rm K^0}^2-m_{\rm K^\pm}^2)^2-4m_{\rm K^0}^2m_{\rm K^\pm}^2}{4s}}
\end{equation}
where
\begin{equation}
s=m_{\rm K^0}^2+m_{\rm K^\pm}^2+2E_{\rm K^0}E_{\rm K^\pm}-2\vec{p}_{\rm K^0}\cdot\vec{p}_{\rm K^\pm}
\end{equation}
and $m_{\rm K^0}$ ($E_{\rm K^0}$) and $m_{\rm K^\pm}$ ($E_{\rm K^\pm}$) are the rest masses
(total energies) of the K$^0_{\rm S}$ and K$^{\rm \pm}$, respectively.

The denominator $B(k^*)$ was formed by mixing K$^0_{\rm S}$ and K$^{\rm \pm}$ particles from each event with K$^{\rm \pm}$ and K$^0_{\rm S}$
particles, respectively, from ten other events, where each event has at least both 
a K$^{\rm \pm}$ and a K$^0_{\rm S}$~\cite{Acharya:2017jks}. 
The vertices of the mixed events were constrained to be within 2 cm of each other in the $z$-direction.

Two-track effects, such as the merging of two real tracks into one reconstructed track and the splitting of one real track into two reconstructed tracks, is an important issue for femtoscopic studies. This analysis dealt with these effects using the following method. For each kaon pair, the distance between the 
K$^0_{\rm S}$ pion daughter track and the same-charged K$^{\rm \pm}$ track was calculated at up to nine points throughout the TPC (every 20 cm from 85 cm to 245 cm) and then averaged. Comparing pairs from the same event to those from mixed events, one observes a splitting peak for an
average separation of $<11$ cm. To correct for this, this analysis demanded that the same-charge particles from each kaon pair must have an average TPC separation of at least 13 cm. Mixed-event tracks were normalized by subtracting the primary vertex position from each used track point.

\begin{figure}
	\centering
		\includegraphics[scale=1.0]{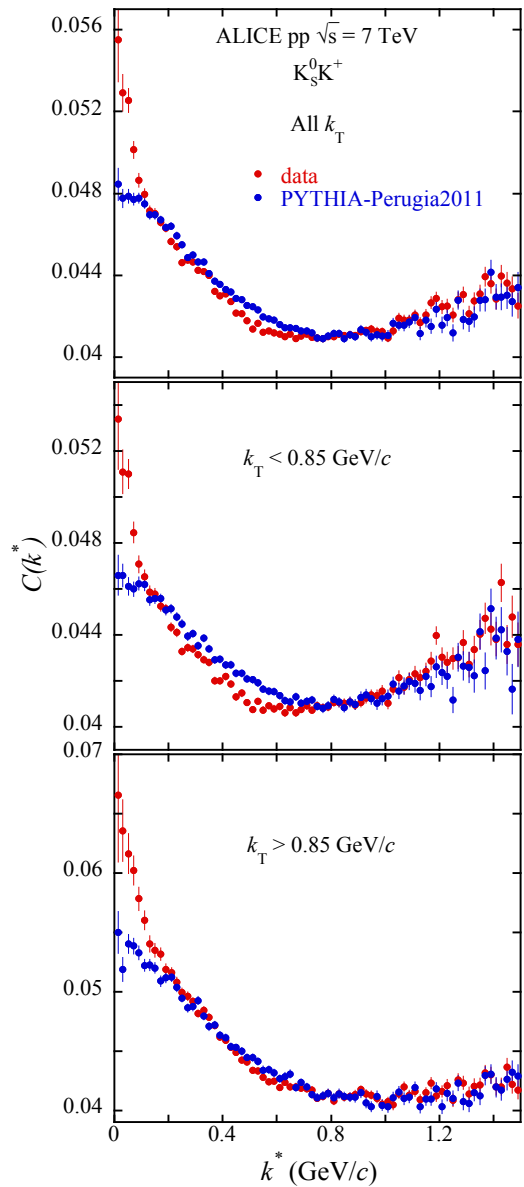}
	\caption{Raw K$^0_{\rm S}$K$^{\rm +}$ correlation functions for the three $k_T$ bins
	compared with those from PYTHIA. The error bars are statistical. The scale of $C(k^*)$ is arbitrary. The PYTHIA correlation functions are normalized to the data
in the vicinity of $k^*=0.8$ GeV/$c$.}
	\label{fig1}
\end{figure}

Correlation functions were created separately for 
the two different charge combinations, K$^0_{\rm S}$K$^{\rm +}$ and K$^0_{\rm S}$K$^{\rm -}$, and for three overlapping/non-exclusive
pair transverse momentum $k_{\rm T} = |\vec{p}_{\rm T,1}+\vec{p}_{\rm T,2}|/2$ ranges: all $k_{\rm T}$, $k_{\rm T}<0.85$ and $k_{\rm T}>0.85$ GeV/$c$, where $k_{\rm T}=0.85$ GeV/$c$ 
is the location of the peak of the $k_{\rm T}$ distribution.
The mean $k_{\rm T}$ values for these three bins were 0.66, 0.49 and 1.17 GeV/$c$, respectively.
The raw K$^0_{\rm S}$K$^{\rm +}$ correlation functions for these three bins compared with those 
generated from PYTHIA simulations with the Perugia2011 tune and
using GEANT3 to model particle transport through the ALICE detectors
are shown in Fig.~\ref{fig1}. The PYTHIA correlation functions are normalized to the data
in the vicinity of $k^*=0.8$ GeV/$c$.
The raw K$^0_{\rm S}$K$^{\rm -}$ correlation functions look very similar to these. 
It is seen that although PYTHIA qualitatively describes the trends of the
baseline of the data, it does not describe it quantitatively such that it could be used to model
the baseline directly.
Instead, for the present analysis the strategy for dealing with the baseline was to describe it
with several functional forms to be fitted to the experimental correlation functions
and to use PYTHIA to test the appropriateness of the proposed baseline functional forms.

\begin{figure}
	\centering
		\includegraphics[scale=.5]{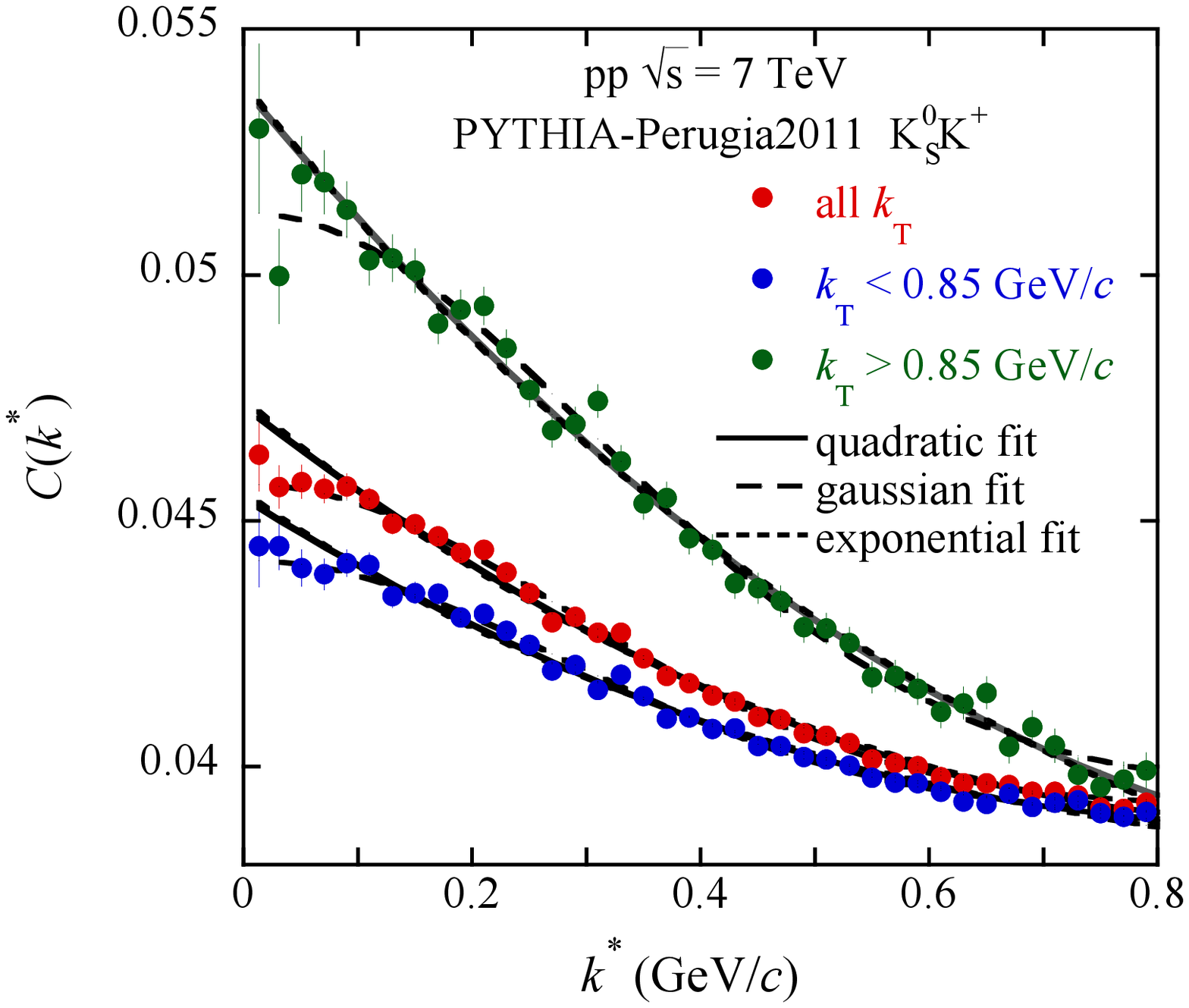}
	\caption{Comparisons of fits of three possible baseline functional forms with the
	PYTHIA correlation functions that were shown in Fig.~\ref{fig1}. Fits were made
	in the $k^*$ range $0 - 0.8$ GeV/$c$. The scale of $C(k^*)$ is arbitrary.}
	\label{fig2}
\end{figure}

Three functional forms for the baseline were tested with PYTHIA: quadratic, Gaussian and 
exponential, given by

\begin{equation}
C_{\rm quadratic}(k^*) = a(1-bk^*+ck^{*2})
\label{quad}
\end{equation}

\begin{equation}
C_{\rm Gaussian}(k^*) = a(1+b\exp(-ck^{*2}))
\label{gauss}
\end{equation}

\begin{equation}
C_{\rm exponential}(k^*) = a(1+b\exp(-ck^*))
\label{exp}
\end{equation}

where $a$, $b$ and $c$ are fit parameters. Fig.~\ref{fig2} shows fits of Eq.~(\ref{quad}),
Eq.~(\ref{gauss}) and Eq.~(\ref{exp}) to the PYTHIA correlation functions shown in
Fig.~\ref{fig1} for the three $k_T$ ranges used in this analysis. As seen, all three functional
forms do reasonably well in representing the PYTHIA correlation functions. Thus, all three
forms were used in fitting the experimental correlation function and the different results
obtained will be used to estimate the systematic uncertainty due to the baseline estimation.
Of course there are an infinite number of functions one could try to represent the baseline,
but at least the three that were chosen for this work are simple and representative of
three basic functional forms.

Correlation functions were corrected for momentum resolution effects using PYTHIA calculations.
The particle momentum resolution in ALICE for the relatively low-momentum tracks used in the
present analysis was $< 10$ MeV/$c$~\cite{Aamodt:2008zz}.
Two correlation functions were generated with PYTHIA: one in terms of the generator-level $k^*$ and one in terms of the simulated detector-level $k^*$. Because PYTHIA does not incorporate final-state interactions, simulated femtoscopic weights were determined using a 9th-order polynomial fit in $k^*$ to the experimental correlation function for the $k_{\rm T}$ range considered.  When filling the same-event distributions, i.e. $A(k^*)$ in Eq.~\ref{Cexp}, kaon pairs were individually weighted by this 9th-order fit according to their generator-level $k^*$. Then, the ratio of the ``ideal'' correlation function to the ``measured'' one (for each $k^*$ bin) was multiplied to the data correlation functions before the fit procedure. This correction mostly affected the lowest $k^*$ bins, increasing the extracted source parameters by $\sim 2\%$.

\subsection{Final-state interaction model}
The final-state interaction model used in the present pp collision analysis follows the same
principles as the ones used for the ALICE Pb-Pb collision analysis~\cite{Acharya:2017jks}.
The measured K$^0_{\rm S}$K$^{\rm \pm}$ correlation functions were fit with formulas that include a parameterization which incorporates strong FSI. It was assumed that the
FSI arises in the K$^0_{\rm S}$K$^{\rm \pm}$ channels due to the near-threshold resonance, $a_0$. This parameterization was introduced by R. Lednicky and is based on the model by R. Lednicky and V.L. Lyuboshitz~\cite{Lednicky:1981su,Lednicky:2005af} (see also Ref.~\cite{Abelev:2006gu} for more details on this parameterization).

Using an equal emission time approximation in the PRF~\cite{Lednicky:1981su}, the 
elastic K$^0_{\rm S}$K$^{\rm \pm}$ transition is written as a stationary solution $\Psi_{-\vkstar}(\vrstar)$ of the scattering problem in the PRF. The quantity
$\vrstar$ represents the emission separation of the pair in the PRF, and
the $-\vkstar$ subscript refers to a reversal of time from the emission process. At large distances this has the asymptotic form of a superposition of an incoming plane wave and an outgoing spherical wave,
\begin{equation}
\Psi_{-\vkstar}(\vrstar) = e^{-i\vkstar \cdot \vrstar} + f(\kstar) \frac{e^{i\kstar\rstar}}{\rstar} \;,
\label{eq:FSIwave}
\end{equation}

where $f(k^*)$ is the $s$-wave K$^0$K$^-$ or ${\rm \overline{K}^0}$K$^+$ scattering amplitude whose contribution is the $s$-wave isovector $a_0$ resonance 
(see Eq.~(11) in Ref.~\cite{Abelev:2006gu}) and

\begin{equation}
f(k^*) = \frac{\gamma_{a_0\rightarrow {\rm K\overline{K}}}}{m_{a_0}^2-s-i(\gamma_{a_0\rightarrow {\rm K\overline{K}}} k^*+\gamma_{a_0\rightarrow \pi\eta}k_{\pi\eta})}\;.
\label{eq:fit4}
\end{equation}

In Eq.~(\ref{eq:fit4}), $m_{a_0}$ is the mass of the $a_0$ resonance, and $\gamma_{a_0\rightarrow {\rm K\overline{K}}}$ 
and $\gamma_{a_0\rightarrow \pi\eta}$ are the couplings of the $a_0$ resonance to the K$^0$K$^-$ (or ${\rm \overline{K}^0}$K$^+$) and $\pi\eta$ channels, respectively. Also, $s=4(m_{\rm K^0}^2+k^{*2})$ and $k_{\pi\eta}$ denotes the momentum in the second decay channel ($\pi\eta$) 
(see Tab.~\ref{table1}).

The correlation function due to the FSI is then calculated by integrating $\Psi_{-\vkstar}(\vrstar)$ in the \textit{Koonin-Pratt equation}~\cite{Koonin:1977fh,Pratt:1990zq},
\begin{equation}
C_{\rm FSI}(\vkstar) = \int {\rm d}^3 \, \vrstar \, S(\vrstar) \left| \Psi_{-\vkstar}(\vrstar) \right| ^2 \, ,
\label{eq:koonin}
\end{equation}
where $S(\vrstar)$ is a one-dimensional Gaussian source function of the PRF relative distance $\left| \vrstar \right| $ with a Gaussian width $R$ of the form
\begin{equation}
S(\vrstar) \sim e^{-\left| \vrstar \right| ^2/(4R^2)}\;.
\label{source}
\end{equation}

Equation~\ref{eq:koonin} can be integrated analytically for K$^0_{\rm S}$K$^{\rm \pm}$ correlations
with FSI for the one-dimensional case, with the result

\begin{equation}
C_{\rm FSI}(k^*)=1+\lambda\alpha\left[\frac{1}{2}\left|\frac{f(k^*)}{R}\right|^2+\frac{2\mathcal{R}f(k^*)}{\sqrt{\pi}R}F_1(2k^* R)-\frac{\mathcal{I}f(k^*)}{R}F_2(2k^* R)+\Delta C\right],
\label{eq:fit2}
\end{equation}
where
\begin{equation}
%F_1(z)\equiv\int^{z}_{0}{\rm d}x\frac{e^{x^2-z^2}}{z};\qquad F_2(z)\equiv\frac{1-e^{-z^2}}{z}.
F_1(z)\equiv\frac{\sqrt{\pi} e^{-z^2} \operatorname{erfi}(z)}{2 z};\qquad F_2(z)\equiv\frac{1-e^{-z^2}}{z}.
\label{eq:fit3}
\end{equation}
In the above equations $\alpha$ is the fraction of K$^{0}_{\rm S}$K$^{\pm}$ pairs that come from the K$^0$K$^-$ or  ${\rm \overline{K}^0}$K$^+$ system, set to 0.5 assuming symmetry in K$^0$ and 
${\rm \overline{K}^0}$ production \cite{Abelev:2006gu}, $R$ is the radius parameter from the
spherical Gaussian source distribution given in Eq.~(\ref{source}), and $\lambda$ is the correlation strength. The correlation strength is
unity in the ideal case of pure $a_0$-resonant FSI, perfect PID, a perfect Gaussian kaon source 
and the absence of long-lived resonances which decay into kaons.
The term $\Delta C$ is a calculated correction factor that takes into account the deviation of the spherical
wave assumption used in Eq.~(\ref{eq:FSIwave}) in the inner region of the short-range potential
(see the Appendix in Ref.~\cite{Abelev:2006gu}). Its effect on the extracted 
$R$ and $\lambda$ parameters
is to increase them by $\sim14\%$. Note that the form of the FSI term in
Eq.~(\ref{eq:fit2}) differs from the form of the FSI term for $\kzs\kzs$ correlations (Eq.~(9) of Ref.~\cite{Abelev:2006gu}) by a factor of $1/2$ due to the non-identical particles in K$^0_{\rm S}$K$^{\rm \pm}$ correlations and thus the absence of the requirement to symmetrize the wavefunction given
in Eq.~(\ref{eq:FSIwave}).

As seen in Eq.~(\ref{eq:fit4}), the K$^0$K$^-$ or ${\rm \overline{K}^0}$K$^+$ s-wave scattering amplitude depends on the $a_0$ mass and decay couplings. From the ALICE 
Pb-Pb collision K$^0_{\rm S}$K$^{\rm \pm}$ study~\cite{Acharya:2017jks}, it was found that source parameters extracted with the ``Achasov2'' parameters of Ref.~\cite{Achasov:2002ir} agreed best with 
the identical kaon measurements, thus in the present pp collision study only the Achasov2
parameters are used. These parameters are shown in Tab.~\ref{table1}. Since the correction
factor $\Delta C$ is found to mainly depend on $\gamma_{a_0K\bar{K}}$~\cite{Abelev:2006gu},
it is judged that the systematic uncertainty on the calculation of $\Delta C$ is negligible. 

\begin{table}
 \centering
 %\begin{tabular}{ | 1 | 1 | 1 | 1 |}
 \begin{tabular}{| c | c | c | c |}
  \hline
  Reference & $m_{a_0}$ & $\gamma_{a_0\rightarrow {\rm K\overline{K}}}$  & $\gamma_{a_0\rightarrow\pi\eta}$ \\ \hline
  %Martin~\cite{Martin1977} & 0.974 & 0.333 & 0.222 \\ \hline
  %Antonelli~\cite{Antonelli2002} & 0.985 & 0.4038 & 0.3711 \\ \hline
  %Achasov1~\cite{Achasov:2002ir} & 0.992 & 0.5555 & 0.4401 \\ \hline
  Achasov2~\cite{Achasov:2002ir} & 1.003 & 0.8365 & 0.4580 \\  
  \hline
  \end{tabular}
  \caption{The $a_0$ mass and coupling parameters, all in GeV/$c^2$, used in
  the present study.}
  \label{table1}
\end{table}

The experimental K$^0_{\rm S}$K$^{\rm \pm}$ correlation functions, calculated using
Eq.~(\ref{Cexp}), were fit with the expression

\begin{equation}
C(k^*) = C_{\rm FSI}(k^*)C_{\rm baseline}(k^*),
\label{eq:corrfit}
\end{equation}

where $C_{\rm baseline}(k^*)$ is Eq.~(\ref{quad}), Eq.~(\ref{gauss}) or Eq.~(\ref{exp}).

The fitting strategy used was to carry out a 5-parameter fit of Eq.~(\ref{eq:corrfit}) to the 
K$^0_{\rm S}$K$^{\rm \pm}$ experimental correlation
functions to extract $R$, $\lambda$, $a$, $b$ and $c$ for each of the six 
($k_T$ range)-(charge state) combinations. For each of these six combinations,
the three baseline functional forms, and two $k^*$ fit ranges, (0.0-0.6 GeV/$c$) and (0.0-0.8 GeV/$c$), were fit, giving six $R$ and six $\lambda$ parameter values for each combination. These six values
were then averaged and the variance calculated to obtain the final values for the parameters
and an estimate of the combined systematic uncertainties from the baseline assumptions and fit
range, respectively. 

\section{Results and discussion}
\subsection{Fits to the experimental correlation functions}
\begin{figure}
	\centering
		\includegraphics[scale=1.4]{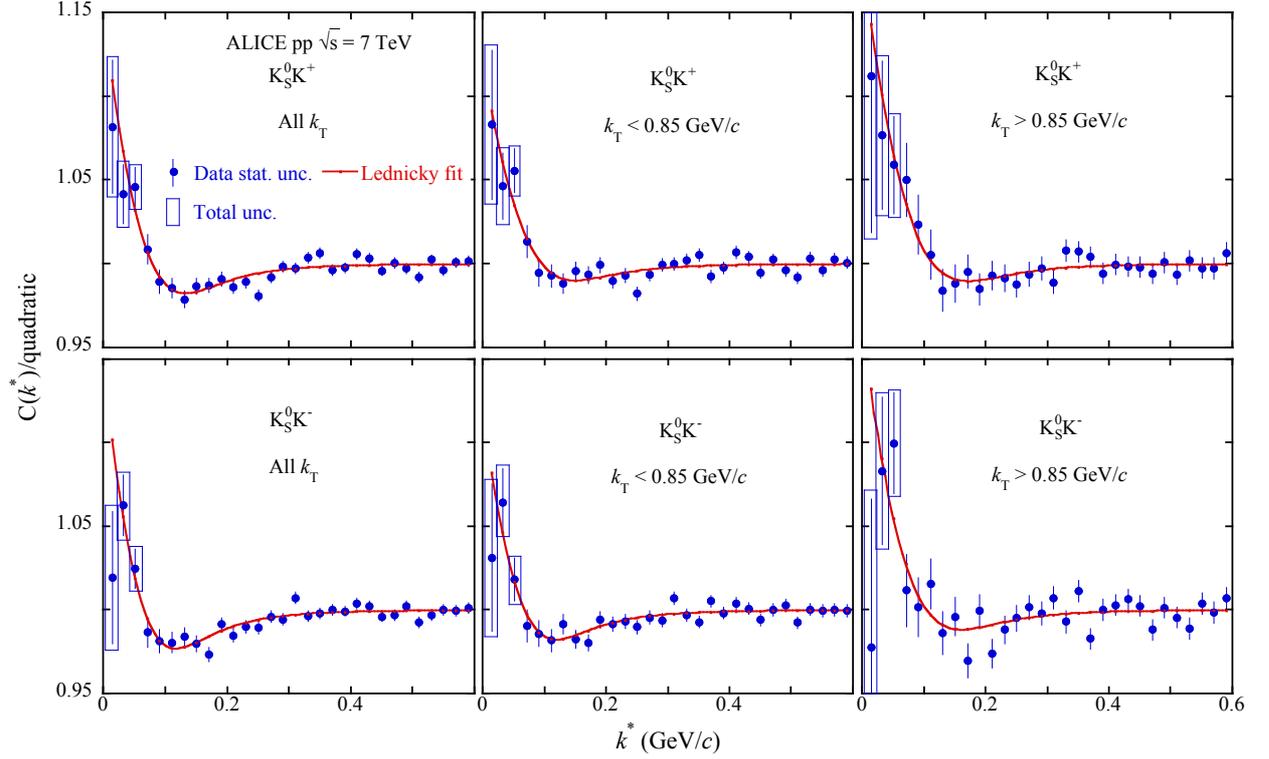}
	\caption{Correlation functions divided by one of the baseline functions with fits from Eq.~(\ref{eq:corrfit}) for 
	K$^0_{\rm S}$K$^+$ and K$^0_{\rm S}$K$^-$ and $k^*$ fit range (0.0-0.6 GeV/$c$) 
	for the three $k_T$ bins and the quadratic baseline function
	assumption. Statistical (lines) and the quadratic sum of the statistical and systematic (boxes) uncertainties are shown. For $k^*>0.05$ GeV/$c$, the systematic uncertainties become negligible and the boxes
	are no longer shown.}
	\label{fig3}
\end{figure}

\begin{figure}
	\centering
		\includegraphics[scale=1.3]{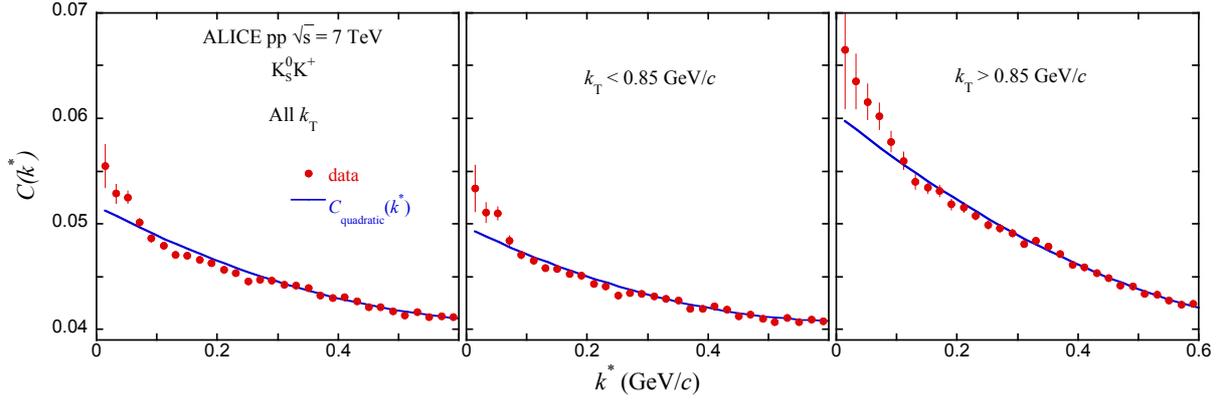}
	\caption{Sample raw correlation functions for K$^0_{\rm S}$K$^+$ 
	showing the fitted quadratic baseline function, Eq.~\ref{quad}.
	Statistical uncertainties are shown. The scale of $C(k^*)$ is arbitrary.}
	\label{fig3a}
\end{figure}

Figure~\ref{fig3} shows sample correlation functions divided 
by the quadratic baseline function with fits of Eq.~(\ref{eq:corrfit}) for 
K$^0_{\rm S}$K$^{\rm \pm}$ and the $k^*$ fit range ($0.0-0.6$ GeV/$c$) for the three $k_T$ bins. The fits using the other baseline assumptions and to the wider range of ($0.0-0.8$ GeV/$c$) are similar in quality. Comparing with the quadratic baseline, using the Gaussian baseline tends to give
$\sim 10-20$\% smaller
source parameters whereas using the exponential baseline tends to give $\sim 10-20$\%
larger source parameters.
The average $\chi^2/{\rm ndf}$ and p-value over all of the fits are 1.554 and 0.172, respectively.
Statistical (lines) and the quadratic sum of the statistical and systematic (boxes) uncertainties are shown.
The systematic uncertainties were determined by varying cuts on the data (See the
discussion of the ``cut systematic uncertainty'' in the section below on ``Systematic Uncertainties
for more details.'').
Fig.~\ref{fig3a} shows sample raw correlation functions for K$^0_{\rm S}$K$^+$ for the three $k_T$ bins and the quadratic baseline function, Eq.~\ref{quad}, that was fit corresponding to the 5-parameter fits of Eq.~\ref{eq:corrfit} 
to the K$^0_{\rm S}$K$^+$ data presented in Fig.~\ref{fig3}.
Statistical uncertainties on the fit parameters were obtained by constructing the 1$\sigma$ $\lambda$ vs. $R$ contour and taking the errors to be at the extreme extents of the contour. A typical value
of the correlation coefficient is 0.642.
This method 
gives the most conservative estimates of the statistical uncertainties. 

The Achasov2 $a_0$ FSI parameterization coupled with the various baseline assumptions gives a good representation of the signal region of the data, i.e. reproducing the enhancement in the $k^*$ region $0.0-0.1$ GeV/$c$ and the small dip in the region $0.1-0.3$ GeV/$c$. A good representation
of the signal region was also seen to be the case for the Pb-Pb analysis with the Achasov2
parameterization, which has a qualitatively different $k^*$ dependence of the correlation
 function that is dominated by a dip at low $k^*$ (Compare present Fig.~\ref{fig3} with Fig.~2 from 
 Ref.~\cite{Acharya:2017jks}).
The enhancement seen for the small-$R$ system at low $k^*$ is expected from Eq.~(\ref{eq:fit2})
as a consequence of the first term in the brackets that goes as $1/R^2$.
This demonstrates the ability of Eq.~(\ref{eq:fit2}) to describe
the FSI in both the small and large size regimes as going through the $a_0$ resonance.

\subsection{Extracted $R$ and $\lambda$ parameters}
The results for the extracted average $R$ and $\lambda$ parameters and the statistical and
systematic uncertainties on these for the present analysis of K$^0_{\rm S}$K$^{\rm \pm}$ femtoscopy
from 7 TeV pp collisions are shown in Tab.~\ref{tab:fitresults}.
The statistical uncertainties given are the averages over the 6 fits for each case.
As can be seen, $R$ and $\lambda$
for K$^0_{\rm S}$K$^+$ agree within the statistical uncertainties with those for K$^0_{\rm S}$K$^-$
in all cases. 

\begin{table}
 \centering
 %\begin{tabular}{ | 1 | 1 | 1 | 1 |}
 \begin{tabular}{| c | c | c | c | c | c | c | c |}
  \hline
\multirow{3}{*}{R or $\lambda$ [+/-]} & $k_{\rm T}$ cut & fit & statistical & fit & cut & total & total \\ 
    & (GeV/$c$)  & value  & uncert.  & systematic  & systematic  & systematic  & quadratic  \\ 
    &     &      &     &  uncert.   & uncert.  &  uncert.   & uncert. \\ \hline
\multirow{3}{*}{R[+] (fm)} &  $k_{\rm T} < 0.85$ & 0.905 & 0.063 & 0.243 & 0.033 & 0.245 & 0.253 \\
    & $k_{\rm T} > 0.85$ & 0.788 & 0.077 & 0.168 & 0.031 & 0.171 & 0.188 \\
    & All $k_{\rm T}$ & 0.922 & 0.048 & 0.188 & 0.038 & 0.192 & 0.198 \\ \hline
\multirow{3}{*}{$\lambda$[+]} & $k_{\rm T} < 0.85$ & 0.189 & 0.046 & 0.070 & 0.012 & 0.071 & 0.085 \\
    & $k_{\rm T} > 0.85$ & 0.222 & 0.080 & 0.066 & 0.015 & 0.068 & 0.105 \\
    & All $k_{\rm T}$ & 0.242 & 0.046 & 0.066 & 0.020 & 0.069 & 0.083 \\ \hline
\multirow{3}{*}{R[-] (fm)} & $k_{\rm T} < 0.85$ & 1.039 & 0.060 & 0.244 & 0.039 & 0.247 & 0.254 \\
    & $k_{\rm T} > 0.85$ & 0.786 & 0.082 & 0.145 & 0.032 & 0.148 & 0.169 \\
    & All $k_{\rm T}$ & 0.995 & 0.046 & 0.185 & 0.041 & 0.190 & 0.195 \\ \hline
\multirow{3}{*}{$\lambda$[-]} & $k_{\rm T} < 0.85$ & 0.253 & 0.044 & 0.096 & 0.016 & 0.097 & 0.107 \\
    & $k_{\rm T} > 0.85$ & 0.208 & 0.084 & 0.038 & 0.016 & 0.042 & 0.094 \\
    & All $k_{\rm T}$ & 0.277 & 0.038 & 0.074 & 0.023 & 0.078 & 0.087 \\
  \hline
  \end{tabular}
  \caption{Fit results for average R and $\lambda$ showing statistical and systematic uncertainties
  from K$^0_{\rm S}$K$^{\rm \pm}$ femtoscopy with pp collisions at $\sqrt{s}=7$ TeV.
  The ``[+/-]'' in the first column refers to K$^0_{\rm S}$K$^+$ or K$^0_{\rm S}$K$^-$. See the text
  for the definitions of the various uncertainties.}
  \label{tab:fitresults}
\end{table}

\subsection{Systematic uncertainties}
Table~\ref{tab:fitresults} shows the total systematic uncertainties on the extracted $R$ and $\lambda$
parameters. As is seen, for most cases the total systematic uncertainty is larger than the statistical
uncertainty. The total systematic uncertainty is broken down in Tab.~\ref{tab:fitresults} into two main contributions,
the ``fit systematic uncertainty'' and the ``cut systematic uncertainty'', and is the quadratic sum
of these. The fit systematic uncertainty is the combined systematic uncertainty due
to the various baseline assumptions and varying the $k^*$ fit range, as described earlier. 
The cut systematic uncertainty is the systematic uncertainty related to the
various cuts made in the data analysis.
To determine this, single particle cuts were varied by $\sim10$\%, and the value chosen for the minimum separation distance of same-sign tracks was varied by $\sim 20$\%. Taking the upper-limit values of the variations to be conservative, this led to additional 
errors of $4\%$ for $R$ and $8\%$ for $\lambda$. As seen in the table, the fit systematic uncertainty dominates
over the cut systematic uncertainty in all cases, demonstrating the large uncertainties in determining the non-femtoscopic baseline in pp collisions. The ``total quadratic uncertainty'' is the
quadratic sum of the ``statistical uncertainty'' column and the ``total systematic uncertainty''
column.

\subsection{Comparisons with K$^0_{\rm S}$K$^{\rm \pm}$ results from Pb-Pb collisions
at $\sqrt{s_{\rm NN}}=2.76$~TeV and identical-kaon results from pp collisions at $\sqrt{s}=7$~TeV}
In this section comparisons of the present results
for $R$ and $\lambda$ with
K$^0_{\rm S}$K$^{\rm \pm}$ measurements from ALICE 2.76~TeV 
Pb-Pb collisions for $0-10$\% centrality~\cite{Acharya:2017jks}, and with
identical-kaon measurements from ALICE 7~TeV pp 
collisions~\cite{Abelev:2012ms,Abelev:2012sq} are presented. Since it is seen in 
Tab.~\ref{tab:fitresults} that the extracted parameters for K$^0_{\rm S}$K$^+$ agree within the statistical uncertainties with those for K$^0_{\rm S}$K$^-$
in all cases, these are averaged over weighted by the statistical uncertainties in the following figures

Figure~\ref{fig4} shows the comparison with the ALICE Pb-Pb collision K$^0_{\rm S}$K$^{\rm \pm}$ measurements. The $\lambda$ parameters have been divided by the pair purity for each case,
i.e. 83\% for the present pp collisions and 88\% for the Pb-Pb collisions~\cite{Acharya:2017jks},
so that they can be compared on the same basis.
It is seen that $R$ for $0-10$\% centrality Pb-Pb is $\sim 5$ fm, and is significantly larger than 
the $R \sim 1$ fm measured for pp collisions. This is expected since $R$ reflects the geometric
size of the interaction region of the collision. It is somewhat surprising that $\lambda$ for pp collisions
is seen to be significantly less than that for Pb-Pb collisions. 
There are two main factors effecting the value of the $\lambda$ parameter: 1) the degree to which a Gaussian fits the correlation function and 2) the effect of long-lived resonances diluting the kaon sample. For 1), it is seen in Fig.~\ref{fig3} for pp and in Fig. 2 of Ref.~\cite{Acharya:2017jks} for Pb-Pb that the Gaussian function used in the Ledincky equation, Eqs.~\ref{source} and \ref{eq:fit2}, fits both colliding systems well, minimizing the effect of 1). For 2), the K$^*$ decay ($\Gamma\sim50$~MeV) has the largest influence on diluting the kaon sample, and it is unlikely that the multiplicity ratio of K/K$^*$ changes dramatically in going from 2.76 TeV to 7 TeV. From these arguments we might naively expect $\lambda$ to be similar in the pp and Pb-Pb cases.

\begin{figure}
	\centering
		\includegraphics[scale=.5]{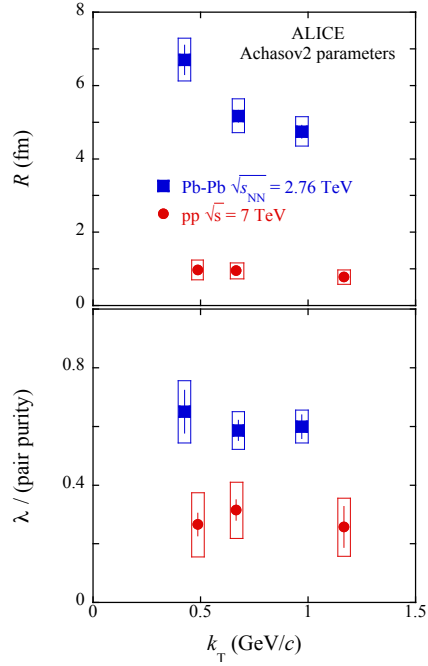}
	\caption{$R$ and $\lambda$ parameters extracted in the present analysis from 
K$^0_{\rm S}$K$^{\rm \pm}$ femtoscopy averaged over K$^0_{\rm S}$K$^+$ and 
K$^0_{\rm S}$K$^-$, along with
a comparison with K$^0_{\rm S}$K$^{\rm \pm}$ results from ALICE 2.76 TeV 
Pb-Pb collisions for $0-10$\% centrality~\cite{Acharya:2017jks}.
The quadratic sum of the statistical and systematic uncertainties is plotted for all results as boxes and
the statistical uncertainties are given as lines. The $\lambda$ parameters have been divided
by their respective pair purities to facilitate their comparison.}
	\label{fig4}
\end{figure}

\begin{figure}
	\centering
		\includegraphics[scale=1.1]{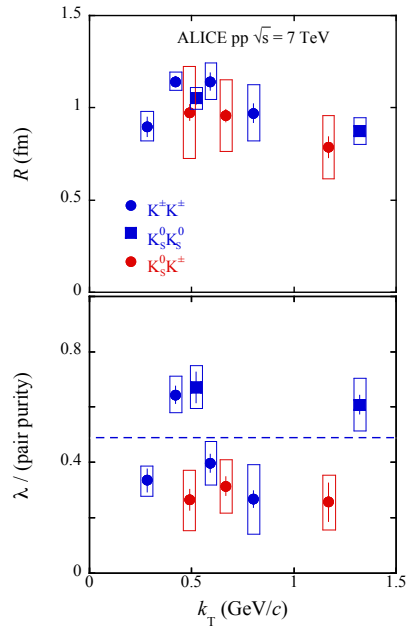}
	\caption{$R$ and purity-normalized $\lambda$ parameters extracted in the present analysis from 
K$^0_{\rm S}$K$^{\rm \pm}$ femtoscopy averaged over K$^0_{\rm S}$K$^+$ and 
K$^0_{\rm S}$K$^-$, along with
comparisons with identical kaon results from ALICE 7 TeV pp collisions averaged over
event multiplicity.
The quadratic sum of the statistical and systematic uncertainties is plotted for all results as boxes and
the statistical uncertainties are given as lines. Also plotted as a blue dashed line
is the simple average of the identical-kaon purity-normalized $\lambda$ parameters.}
	\label{fig5}
\end{figure}

In order to properly compare the present results with the ALICE pp collision identical-kaon 
measurements, we must
take the weighted average (weighted by their statistical uncertainties) over the multiplicity bins used in 
Refs.~\cite{Abelev:2012ms,Abelev:2012sq} since our present results are summed over all multiplicity.
Figure~\ref{fig5} shows the comparison between the present results for $R$ and $\lambda$
and measurements from the identical-kaon femtoscopy in 7 TeV pp collisions. The $R$ values
are seen to agree between the present analysis and the identical kaon analyses within the uncertainties. The $\lambda$ parameters shown in Fig.~\ref{fig5} are each divided by 
their respective pair 
purities. Going from the lowest to the highest $k_{\rm T}$ points, for the neutral-kaon pairs
the purities are 0.88 and 0.84~\cite{Abelev:2012ms}, and
for the charge-kaon pairs the purities are 0.84, 0.61, 0.79 and 1.0~\cite{Abelev:2012sq}, respectively.
The purity-normalized $\lambda$ parameters for the identical kaons are seen to scatter in a wide range between values of $0.3 - 0.7$, 
whereas the K$^0_{\rm S}$K$^{\rm \pm}$
values are seen to lie in the narrower range of $0.25 - 0.30$.

In order to help to clarify the comparison between the purity-normalized $\lambda$ values from K$^0_{\rm S}$K$^{\rm \pm}$ and the identical-kaon results,  the simple average 
over the identical kaon purity-normalized $\lambda$ parameters is plotted as a blue dashed line
in Fig.~\ref{fig5}. As seen, the K$^0_{\rm S}$K$^{\rm \pm}$
values tend to be smaller than the average of the identical kaons, as was more significantly
the case for the comparison with the purity-normalized $\lambda$ values from Pb-Pb seen in Fig.~\ref{fig4},
however the large scatter of the identical kaons makes it difficult to draw any
strong conclusions from this comparison.

\subsection{Implications from the present results for the $a_0$ to be a tetraquark state}
The K$^0_{\rm S}$K$^{\rm \pm}$ FSI is described
well by assuming it is due to the $a_0$ resonance for both pp and Pb-Pb collisions,
as seen in Fig.~\ref{fig3} of the present work and in Fig.~2 of Ref.~\cite{Acharya:2017jks}.
The $R$ parameters extracted from this method are also seen to agree within uncertainties
with the identical-kaon measurements for each of these collision systems. For Pb-Pb
collisions, it was found that the $\lambda$ parameters extracted from K$^0_{\rm S}$K$^{\rm \pm}$
also agree with the corresponding identical-kaon measurements for Pb-Pb collisions indicating that
the FSI between the kaons goes solely through the $a_0$ resonance.
The present pp collision results for $\lambda$, which are significantly
lower than the K$^0_{\rm S}$K$^{\rm \pm}$ values from 
Pb-Pb collisions seen in Fig.~\ref{fig4} and which tend to be
lower than the corresponding identical-kaon
values in pp collisions seen in Fig.~\ref{fig5},
imply that the FSI for these collisions does not go solely
through the $a_0$ resonance, i.e. non-resonant elastic channels and/or free-streaming
are also present. From the arguments given in the Introduction, this is the geometric effect
that would be expected
in the case of a tetraquark $a_0$ since competing annihilation channels could open up
in the smaller system and compete with the FSI through the $a_0$, whereas for a diquark
$a_0$ the FSI should still go solely through the $a_0$.
The pp collision results are thus compatible with
the conclusion from the Pb-Pb collision measurement~\cite{Acharya:2017jks} that favors the
interpretation of the $a_0$ resonance to be a tetraquark state.

\section{Summary}
In summary, femtoscopic correlations with the particle pair combinations K$^0_{\rm S}$K$^{\rm \pm}$
are studied in pp collisions at $\sqrt{s}=7$ TeV for the first time by the LHC ALICE experiment. Correlations in the K$^0_{\rm S}$K$^{\rm \pm}$ pairs are produced by final-state
interactions which proceed through the $a_0$ resonance. It is found that the $a_0$
final-state interaction describes the shape of the measured K$^0_{\rm S}$K$^{\rm \pm}$ correlation
functions well. The extracted radius and $\lambda$ parameters for K$^0_{\rm S}$K$^{\rm -}$ are found to be equal
within the experimental uncertainties to those for K$^0_{\rm S}$K$^{\rm +}$. Results of the present study are compared with those
from identical-kaon femtoscopic studies also performed with pp collisions at $\sqrt{s}=7$ TeV by ALICE
and with a recent ALICE K$^0_{\rm S}$K$^{\rm \pm}$ measurement in Pb-Pb collisions at $\sqrt{s_{\rm NN}}=2.76$ TeV.
These comparisons suggest that non-resonant elastic scattering channels are present in pp collisions,
unlike in Pb-Pb collisions. It is our conclusion that the present results, in combination 
with the ALICE Pb-Pb collision measurements, favor the interpretation of the $a_0$ to be a tetraquark state.

%%%%% acknowledgements
\newenvironment{acknowledgement}{\relax}{\relax}
\begin{acknowledgement}
\section*{Acknowledgements}
% Version: 2018-07-30

The ALICE Collaboration would like to thank all its engineers and technicians for their invaluable contributions to the construction of the experiment and the CERN accelerator teams for the outstanding performance of the LHC complex.
The ALICE Collaboration gratefully acknowledges the resources and support provided by all Grid centres and the Worldwide LHC Computing Grid (WLCG) collaboration.
The ALICE Collaboration acknowledges the following funding agencies for their support in building and running the ALICE detector:
A. I. Alikhanyan National Science Laboratory (Yerevan Physics Institute) Foundation (ANSL), State Committee of Science and World Federation of Scientists (WFS), Armenia;
Austrian Academy of Sciences and Nationalstiftung f\"{u}r Forschung, Technologie und Entwicklung, Austria;
Ministry of Communications and High Technologies, National Nuclear Research Center, Azerbaijan;
Conselho Nacional de Desenvolvimento Cient\'{\i}fico e Tecnol\'{o}gico (CNPq), Universidade Federal do Rio Grande do Sul (UFRGS), Financiadora de Estudos e Projetos (Finep) and Funda\c{c}\~{a}o de Amparo \`{a} Pesquisa do Estado de S\~{a}o Paulo (FAPESP), Brazil;
Ministry of Science \& Technology of China (MSTC), National Natural Science Foundation of China (NSFC) and Ministry of Education of China (MOEC) , China;
Ministry of Science and Education, Croatia;
Centro de Aplicaciones Tecnol\'{o}gicas y Desarrollo Nuclear (CEADEN), Cubaenerg\'{\i}a, Cuba;
Ministry of Education, Youth and Sports of the Czech Republic, Czech Republic;
The Danish Council for Independent Research | Natural Sciences, the Carlsberg Foundation and Danish National Research Foundation (DNRF), Denmark;
Helsinki Institute of Physics (HIP), Finland;
Commissariat \`{a} l'Energie Atomique (CEA) and Institut National de Physique Nucl\'{e}aire et de Physique des Particules (IN2P3) and Centre National de la Recherche Scientifique (CNRS), France;
Bundesministerium f\"{u}r Bildung, Wissenschaft, Forschung und Technologie (BMBF) and GSI Helmholtzzentrum f\"{u}r Schwerionenforschung GmbH, Germany;
General Secretariat for Research and Technology, Ministry of Education, Research and Religions, Greece;
National Research, Development and Innovation Office, Hungary;
Department of Atomic Energy Government of India (DAE), Department of Science and Technology, Government of India (DST), University Grants Commission, Government of India (UGC) and Council of Scientific and Industrial Research (CSIR), India;
Indonesian Institute of Science, Indonesia;
Centro Fermi - Museo Storico della Fisica e Centro Studi e Ricerche Enrico Fermi and Istituto Nazionale di Fisica Nucleare (INFN), Italy;
Institute for Innovative Science and Technology , Nagasaki Institute of Applied Science (IIST), Japan Society for the Promotion of Science (JSPS) KAKENHI and Japanese Ministry of Education, Culture, Sports, Science and Technology (MEXT), Japan;
Consejo Nacional de Ciencia (CONACYT) y Tecnolog\'{i}a, through Fondo de Cooperaci\'{o}n Internacional en Ciencia y Tecnolog\'{i}a (FONCICYT) and Direcci\'{o}n General de Asuntos del Personal Academico (DGAPA), Mexico;
Nederlandse Organisatie voor Wetenschappelijk Onderzoek (NWO), Netherlands;
The Research Council of Norway, Norway;
Commission on Science and Technology for Sustainable Development in the South (COMSATS), Pakistan;
Pontificia Universidad Cat\'{o}lica del Per\'{u}, Peru;
Ministry of Science and Higher Education and National Science Centre, Poland;
Korea Institute of Science and Technology Information and National Research Foundation of Korea (NRF), Republic of Korea;
Ministry of Education and Scientific Research, Institute of Atomic Physics and Romanian National Agency for Science, Technology and Innovation, Romania;
Joint Institute for Nuclear Research (JINR), Ministry of Education and Science of the Russian Federation and National Research Centre Kurchatov Institute, Russia;
Ministry of Education, Science, Research and Sport of the Slovak Republic, Slovakia;
National Research Foundation of South Africa, South Africa;
Swedish Research Council (VR) and Knut \& Alice Wallenberg Foundation (KAW), Sweden;
European Organization for Nuclear Research, Switzerland;
National Science and Technology Development Agency (NSDTA), Suranaree University of Technology (SUT) and Office of the Higher Education Commission under NRU project of Thailand, Thailand;
Turkish Atomic Energy Agency (TAEK), Turkey;
National Academy of  Sciences of Ukraine, Ukraine;
Science and Technology Facilities Council (STFC), United Kingdom;
National Science Foundation of the United States of America (NSF) and United States Department of Energy, Office of Nuclear Physics (DOE NP), United States of America.    %%%%%%% done by webmaster team
\end{acknowledgement}

%%%%%%%% Bibliography (In case of using bibtex generate the bbl requested by arXiv)
\bibliographystyle{utphys}   % Remember we use title in the biblio
\bibliography{K0sKchpp_lettervx.bib}
%\input {K0sKch_lettervx.bib}  

%%%%%%%%% appendix with author list
\newpage
\appendix
\section{The ALICE Collaboration}
\label{app:collab}
% Collaboration: CERN-LHC-ALICE
% Generation Date is 2018-09-21 (manual generation)

% How to use:
%%%%%%%%% appendix with author list
%\appendix
%\section{The ALICE Collaboration}
%\label{app:collab}
%\input{authors-list.tex}  %%%%%%% get the latest version before submitting

\begingroup
\small
\begin{flushleft}

S.~Acharya$^{\rm 139}$, 
F.T.-.~Acosta$^{\rm 20}$, 
D.~Adamov\'{a}$^{\rm 93}$, 
A.~Adler$^{\rm 74}$, 
J.~Adolfsson$^{\rm 80}$, 
M.M.~Aggarwal$^{\rm 98}$, 
G.~Aglieri Rinella$^{\rm 34}$, 
M.~Agnello$^{\rm 31}$, 
N.~Agrawal$^{\rm 48}$, 
Z.~Ahammed$^{\rm 139}$, 
S.U.~Ahn$^{\rm 76}$
% ...

\bigskip

\bigskip 

\textbf{\Large Affiliation Notes}

\bigskip 

% ...
$^{\rm V}$ Also at: Institute of Theoretical Physics, University of Wroclaw, Poland\\

\bigskip

\bigskip 

\textbf{\Large Collaboration Institutes}

\bigskip 

% ...
$^{20}$ Department of Physics, University of California, Berkeley, California, United States\\
% ...
$^{31}$ Dipartimento DISAT del Politecnico and Sezione INFN, Turin, Italy\\
% ...
$^{34}$ European Organization for Nuclear Research (CERN), Geneva, Switzerland\\
% ...
$^{48}$ Indian Institute of Technology Bombay (IIT), Mumbai, India\\
% ...
$^{74}$ Johann-Wolfgang-Goethe Universit\"{a}t Frankfurt Institut f\"{u}r Informatik, Fachbereich Informatik und Mathematik, Frankfurt, Germany\\
% ...
$^{76}$ Korea Institute of Science and Technology Information, Daejeon, Republic of Korea\\
% ...
$^{80}$ Lund University Department of Physics, Division of Particle Physics, Lund, Sweden\\
% ...
$^{93}$ Nuclear Physics Institute of the Czech Academy of Sciences, \v{R}e\v{z} u Prahy, Czech Republic\\
% ...
$^{98}$ Physics Department, Panjab University, Chandigarh, India\\
% ...
$^{139}$ Variable Energy Cyclotron Centre, Homi Bhabha National Institute, Kolkata, India\\
% ...

\endgroup  %%%%%%% done by webmaster team
\end{document}